\documentclass{SciPost}

\hypersetup{
    colorlinks,
    linkcolor={red!50!black},
    citecolor={blue!50!black},
    urlcolor={blue!80!black}
}

\DeclareSymbolFont{usualmathcal}{OMS}{cmsy}{m}{n}
\DeclareSymbolFontAlphabet{\mathcal}{usualmathcal}

\fancypagestyle{SPstyle}{
\fancyhf{}
\lhead{\colorbox{scipostblue}{\bf \color{white} ~SciPost Physics }}
\rhead{{\bf \color{scipostdeepblue} ~Submission }}

\fancyfoot[C]{\textbf{\thepage}}
}

\usepackage{dcolumn}
\usepackage{bm}
\usepackage{physics}
\usepackage{url}

\usepackage{tikz}
\usepackage{tikzscale}
\usetikzlibrary{decorations.pathmorphing}
\usepackage{pgfplots}
\pgfplotsset{compat=1.9}
\usepackage{algorithm}
\usepackage{algpseudocode}
\usepackage{bbold}
\usepgfplotslibrary{statistics}
\usetikzlibrary{shapes, arrows.meta}

\usepackage{xspace}

\usepackage{multirow}

\usepackage{subcaption}
\usepackage{float}

\usepackage[capitalise, noabbrev]{cleveref}

\usepackage{units}

\newcommand{\geant}{\textsc{Geant4}\xspace}
\newcommand{\cpf}{\textsc{CaloPointFlow}\xspace}
\newcommand{\cpfii}{\textsc{CaloPointFlow II}\xspace}
\newcommand{\deepsetflow}{\textsc{DeepSetFlow}\xspace}
\newcommand{\cdfdequantization}{\textsc{CDF-Dequantization}\xspace}
\newcommand{\challenge}{\textsc{CaloChallenge}\xspace}

\begin{document}

\pagestyle{SPstyle}

\begin{center}{\Large \textbf{\color{scipostdeepblue}{
\cpfii \\ Generating Calorimeter Showers as Point Clouds
}}}\end{center}

\begin{center}\textbf{
Simon Schnake\textsuperscript{1,2$\star$},
Dirk Krücker\textsuperscript{1} and
Kerstin Borras\textsuperscript{1,2}
}\end{center}

\begin{center}
{\bf 1} Deutsches Elektronen-Synchrotron DESY, Germany
\\
{\bf 2} RWTH Aachen University, Germany
\\[\baselineskip]
$\star$ \href{mailto:simon.schnake@desy.de}{\small simon.schnake@desy.de}
\end{center}

\section*{\color{scipostdeepblue}{Abstract}}
\textbf{%
The simulation of calorimeter showers presents a significant computational challenge, impacting the efficiency and accuracy of particle physics experiments. While generative ML models have been effective in enhancing and accelerating the conventional physics simulation processes, their application has predominantly been constrained to fixed detector readout geometries. With \cpf we have presented one of the first models that can generate a calorimeter shower as a point cloud.  This study describes \cpfii, which exhibits several significant improvements compared to its predecessor. This includes a novel dequantization technique, referred to as \cdfdequantization, and a normalizing flow architecture, referred to as \deepsetflow. The new model was evaluated with the fast Calorimeter Simulation Challenge (\challenge) Dataset II and III.

}

\vspace{\baselineskip}


\vspace{10pt}
\noindent\rule{\textwidth}{1pt}
\tableofcontents
\noindent\rule{\textwidth}{1pt}
\vspace{10pt}



\section{Introduction}

Accurately simulating particle interactions in detectors is a fundamental part of modern high-energy physics (HEP) experiments, such as those performed at the Large Hadron Collider (LHC). These simulations are essential for physicists to produce data that can be compared with experimental results, aiding in the search for new physics phenomena or in making more precise measurements of known physical properties. The planned high luminosity upgrade of the LHC \cite{ZurbanoFernandez:2020cco} will see experimental data taken at more and more increasing rates, presenting significant challenges for coping with the simulation needs. Among the various components of a detector, calorimeters are especially demanding in simulation. Calorimeters are used to measure the energy of incoming particles by detecting the cascade of secondary particles they produce. To simulate these processes, detailed and complex multistep computations are necessary.

Conventionally, simulations such as \geant \cite{geant_simul_toolkit} are employed to accurately replicate these intricate interactions. While these methods offer unmatched accuracy, they are computationally intensive, often requiring seconds per event on a conventional CPU. This high computational demand creates a bottleneck, especially as we move into an era of high-luminosity experiments that will produce larger volumes of data, feature more complex detector geometries, and demand simulations of unprecedented scale and quality. Moreover, the projected computing budgets at large experiments are difficult to reconcile with the increasing amount of simulated events needed, given the current capabilities of Monte Carlo simulations \cite{HEPSoftwareFoundation:2017ggl, Boehnlein:2803119}. The detailed simulations consume a significant portion of the computational budget in many large-scale experiments, further sharpening the challenge.

Given these constraints, it is increasingly impossible to run full, detailed detector simulations for each event to be simulated. The use of \emph{fast simulation} methods is on the rise, as they approximate high-fidelity simulations while occupying less computational power. However, conventional fast simulation frameworks, which are mostly based on parametric models \cite{Barberio:2009zza, Barberio:2008zza, Rahmat:2011xp, Rahmat:2013sva, Rahmat:2014tza, Jang:2009zua, Jun:2011zz, Grindhammer:1993kw, Mahboubi:2005wya, Grindhammer:1989zg}, often fail to capture subtle details of calorimeter interactions. This lack of precision can lead to discrepancies in results when performing physics analyses, which emphasizes the need for a more efficient, yet accurate, alternative.

To tackle these challenges, there is an increasing effort to develop generative machine learning models as a more computationally efficient yet accurate method for calorimeter simulation. The objective is to develop models that can precisely reproduce the distribution of calorimeter responses while significantly reducing computational time and resources. Such an approach not only reduces the burden on the computational infrastructure, but also has the potential to enable more intricate and precise analyses than those currently feasible.

To address these aspects, there is a growing field of research about deep learning based surrogate models \cite{Paganini:2017hrr, Paganini:2017dwg, deOliveira:2017rwa, Erdmann:2018kuh, Erdmann:2018jxd, Belayneh:2019vyx, buhmann2020getting, ATL-SOFT-PUB-2018-001, ATLAS:2022jhk, Krause:2021ilc, Krause:2021wez, Buhmann:2021caf, ATLAS:2021pzo, Mikuni:2022xry, Mikuni:2023tqg, ATLAS:2022jhk, Adelmann:2022ozp, Krause:2022jna, Cresswell:2022tof, AbhishekAbhishek:2022wby, schnakegenerating, Diefenbacher:2023vsw, Diefenbacher:2023prl, Buhmann:2023bwk, Buhmann:2023kdg, Hashemi:2023ruu, Dubinski:2023fsy, Acosta:2023zik, Amram:2023onf, Pang:2023wfx, Hashemi:2023rgo, Ernst:2023qvn, Buckley:2023rez, Diefenbacher:2023flw, Erdmann:2023ngr, kach:2023neurips, Kobylianskii:2024ijw} such as Generative Adversarial Networks (GANs) \cite{Paganini:2017hrr, Paganini:2017dwg, deOliveira:2017rwa, Erdmann:2018kuh, Erdmann:2018jxd, Belayneh:2019vyx, buhmann2020getting, ATL-SOFT-PUB-2018-001, ATLAS:2022jhk, Buhmann:2021caf, ATLAS:2021pzo, ATLAS:2022jhk, Diefenbacher:2023prl, Hashemi:2023ruu, Dubinski:2023fsy, Erdmann:2023ngr, kach:2023neurips}, Variational Autoencoders (VAEs) \cite{buhmann2020getting, Buhmann:2021caf, AbhishekAbhishek:2022wby, Ernst:2023qvn}, normalizing flows \cite{Krause:2021ilc, Krause:2021wez, Krause:2022jna, schnakegenerating, Diefenbacher:2023vsw, Pang:2023wfx, Ernst:2023qvn, Buckley:2023rez}, and diffusion models \cite{Mikuni:2022xry, Mikuni:2023tqg, Buhmann:2023bwk, Buhmann:2023kdg, Acosta:2023zik, Amram:2023onf, Diefenbacher:2023flw, Kobylianskii:2024ijw}, that have been developed for detector simulations. 

These models replicate the output of traditional simulations, such as \geant, and are designed to emulate the complex interactions of particles within calorimeters. This is achieved with significantly fewer computational resources. This progress is particularly important in high-energy physics (HEP) experiments, as it not only mitigates computational bottlenecks but also enables more comprehensive and detailed investigations.

Most of these models are based on a fixed data geometry, where a calorimeter is represented by a collection of voxels. Each voxel corresponds to a single calorimeter sensor. High granular calorimeters are composed of several million sensor cells. For example, the proposed upgrade for the CMS Calorimeter, known as the HGCAL \cite{CMS:2017jpq}, is expected to feature around 6 million sensor cells. However, it is commonly observed that particle showers deposit their energy in only a small portion of the total number of cells, resulting in a sparsely populated voxel representation.

Therefore, it is more efficient to model the distribution of hits, which represent the locations where energy is actually deposited, rather than attempting to represent every single cell. These 'hits' can be conceptualized as points in a four-dimensional space, combining three spatial dimensions with an additional energy component. The number of points detected by the calorimeter corresponds to the total number of hits.

This approach is consistent with previous studies in particle physics that have investigated generative models based on point clouds \cite{Buhmann:2023pmh, Buhmann:2023bwk, Buhmann:2023kdg, Kach:2022uzq, Kach:2023rqw, Mikuni:2023dvk, Scham:2023usu, Scham:2023cwn, Kansal:2021cqp, Buhmann:2023zgc, Leigh:2023toe, Leigh:2023zle, birk2023flow, Acosta:2023zik, kach:2023neurips, schnakegenerating}. Previous research has also explored the use of these models for calorimeter simulations \cite{schnakegenerating, Buhmann:2023bwk, Buhmann:2023kdg, kach:2023neurips, Acosta:2023zik}.

We have developed \cpf, one of the first point cloud-based surrogate models specifically designed for calorimeter simulation \cite{schnakegenerating}. Building on this model, we present here the advanced version of the original model, which we have named \cpfii\texttt{\footnote{The code for this study can be found at \href{https://github.com/simonschnake/CaloPointFlow}{github.com/simonschnake/CaloPointFlow}.}}. \cpfii has been tailored for the Fast Calorimeter Simulation Challenge (\challenge) \cite{CaloChallenge}, and incorporates several significant updates compared to the original \cpf architecture. They are described below:

\begin{enumerate}
\item A major boost in \cpfii is the introduction of a novel dequantization technique, called \cdfdequantization. This technique is a solution to the common problems in dequantization, ensuring that the marginal distributions are normally distributed.
\item Another major development in \cpfii is the creation of a new Normalizing Flow architecture, referred to as \deepsetflow. This architecture allows the modeling of point-to-point correlations. Such correlations were difficult to capture in the previous model, but \deepsetflow overcomes this limitation.
\item In addition, \cpfii leverages the rotational symmetry inherent in the datasets. This method mitigates the "multiple hit" problem, which has been a notable obstacle in accurately simulating the behavior of showers within calorimeters.
\end{enumerate}

Our paper is structured as follows. First, we describe the datasets used in \cref{sec:datasets}. Then we describe our model in \cref{sec:model} and briefly line out our pre- and post-processing in \cref{sec:prepostprocessing}.
Next, we introduce \cdfdequantization in \cref{sec:cdfdequantization}. In \cref{sec:deepsetflow} we explain details of   \deepsetflow and in \cref{sec:multiplehitkludge}, we describe our mitigation strategy for the multiple hit problem. In \cref{sec:results} we present the results and evaluate the model. Finally, we summarize the paper with conculsions in \cref{sec:conclusion}. 

\section{Datasets}
\label{sec:datasets}

In our research, we exclusively use the second and third dataset from the \challenge \cite{CaloChallenge}. Each shower in the dataset contains the incident energy of the incoming particle and a vector containing the voxel energies.  The incident energy $E_{\rm inc}$ is log-uniformly distributed between $1 \unit[1]{GeV}$ and $1 \unit[1]{TeV}$. Each dataset consists of 200,000 showers initiated by electrons. These datasets are equally divided for training and evaluation purposes. 

Datasets 2 \cite{giannelli:2022ccd2} and Dataset 3 \cite{giannelli:2022ccd3} are simulated using the same physical detector, which consists of concentric cylinders with 90 layers of absorber and sensitive (active) materials, specifically tungsten (W) and silicon (Si), respectively. Each sub-layer consists of $1.4 \unit[1]{mm}$ of W and $0.3 \unit[1]{mm}$ of Si, resulting in a total detector depth of $153\unit[1]{mm}$. The detector's inner radius is $80 \unit[1]{cm}$.

The readout segmentation is determined by the direction of the particle entering the calorimeter. This direction defines the z-axis of the cylindrical coordinate system, with the entrance position in the calorimeter set as the origin $(0,0,0)$. The voxels (readout cells) in both datasets 2 and 3 have identical sizes along the z-axis but differ in segmentation in radius (r) and angle ($\alpha$).

For the $z$-axis, the voxel size is $3.4\mathrm{mm}$, corresponding to two physical layers (W-Si-W-Si). Considering only the absorber value of the radiation length $\unit[1]{X_{0}}(W) = 3.504 \unit[1]{mm}$, the $z$-cell size equates to $2 \times \frac{1.4 \unit[1]{mm}}{3.504 \unit[1]{mm}} = 0.8 \unit[1]{X_{0}}$. In the radial dimension, the cell sizes are $2.325 \unit[1]{mm}$ for dataset 3 and $4.65 \unit[1]{mm}$ for dataset 2. Approximately, considering the Molière radius of Tungsten (W) only, this corresponds to $0.25\unit[1]{R_m}$ for dataset 3 and $0.5\unit[1]{R_m}$ for dataset 2. The minimum energy threshold for the readout per voxel in datasets 2 and 3 is set to $15.15 \mathrm{keV}$.

The calorimeter geometry of Dataset 2 comprises 45 concentric cylindrical layers stacked along the direction of particle propagation ($z$). Each layer is further divided into 16 angular bins ($\alpha$) and nine radial bins ($r$), resulting in a total of $45\times 16 \times 9 = 6480$ voxels.

Dataset 3 from the \challenge features a higher granularity compared to Dataset 2. 
Each layer in Dataset 3 consists of 18 radial bins and 50 angular bins, resulting in a total of $45\times50\times18 = 40500$ voxels. 

For both datasets, we adhered to the \challenge's specifications by dividing the available events equally between training and evaluation.

\section{Model}
\label{sec:model}

\begin{figure}[H]
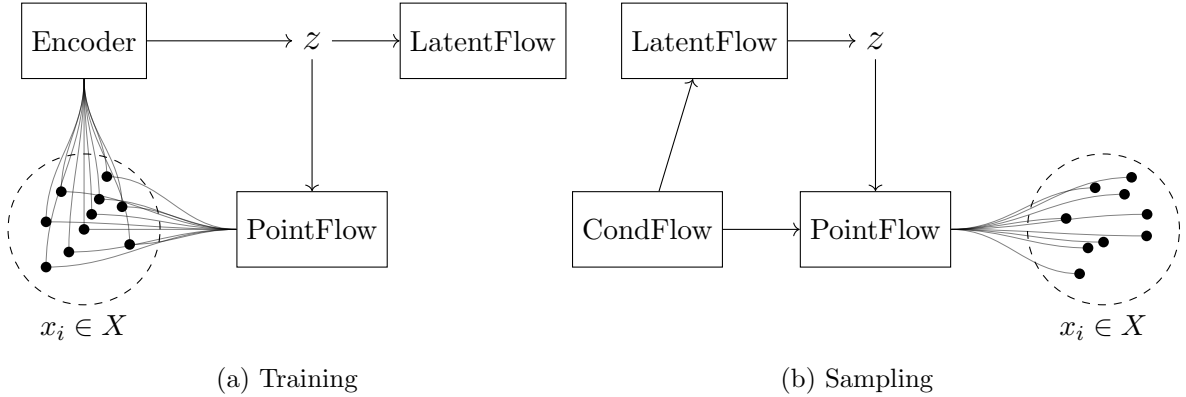

    \centering
    \begin{subfigure}{.5\textwidth}
    \centering
    \includegraphics{visualizations/model_training.tikz}
    \caption{Training}
    \label{fig:model-training}
    \end{subfigure}%
    \begin{subfigure}{.5\textwidth}
    \centering
    \includegraphics{visualizations/model_generating.tikz}
    \caption{Sampling}
    \label{fig:model-sampling}
    \end{subfigure}%
    \caption{A schematic view of the model. Part a) displays the training setup. The \emph{Encoder} encodes the points into the latent representation $z$. The \emph{LatentFlow} is optimized to learn $z$, while the \emph{PointFlow} is optimized to learn the distribution of points $x_i$ conditioned on $z$. Part b) displays the sampling setup. First, the conditional variables $E_{\text{sum}}$ and $n_{\text{hits}}$ are generated. Then, the \emph{LatentFlow} generates the latent representation $z$, which is used to generate the points with the \emph{PointFlow}.}rar
\end{figure}

The model presented in this section is a continued development from our previous model \cite{schnakegenerating}. It is evolved from the basics of the \emph{PointFlow} model, which was introduced by Yang et al. \cite{Yang:2019pointflow}.

The target of the model is to generate point clouds with the probability density $p(X)$, where $X$ are the possible showers. The schematic layout of the data flow in the training process can be seen in \cref{fig:model-training}.
The model is considered a \emph{variational autoencoder} (VAE) \cite{vae}, with the \emph{PointFlow} serving as the decoder and the \emph{LatentFlow} transforming the encoded distribution. To train the model, the Evidence Lower Bound (ELBO) is maximized. The encoder $q_\varphi(z | X)$ and decoder $p_\theta(X|z)$ can be trained simultaneously over the probability distribution of the latent variable $p(z)$ with this loss-function

\begin{align}
    \mathcal{L}  &= \mathbb{E}_{q_\varphi(z | X)}[ \ln p_\theta(X|z, \mathcal{C})] - D_{KL}(q_\varphi(z|X) || p(z| \mathcal{C}))\\ 
    &= \underbrace{\mathbb{E}_{q_\varphi(z | X)}\left[\ln p_\theta(X|z, \mathcal{C}) \right]}_{\mathcal{L}_{\textrm{recon}}} + \underbrace{\mathbb{E}_{q_\varphi(z | X)}\left[\ln p_\theta(z | \mathcal{C}) \right] }_{\mathcal{L}_{\textrm{prior}}} - \mathcal{H}(q_\varphi(z|X)) .
\end{align}

The loss-function consists of three parts. The first part $\mathcal{L}_{\textrm{recon}}$ is the reconstruction error of the shower $X$. Minimizing $\mathcal{L}_{\textrm{recon}}$ is equal to generating hits of the shower with a high likelihood. The second part $\mathcal{L}_{\textrm{prior}}$ is the expectation value of the prior distribution obeying the approximated distribution of the encoder.
Minimizing $\mathcal{L}_{\textrm{prior}}$  is equal to generating a latent vector $z$ with a high probability. The last term $\mathcal{H}(q_\varphi(z|X))$ is the entropy of the encoded values. It is a regularization on the latent space $z$.

Here the encoder $q_\varphi(z|x)$ approximates $z$ conditioned on $X$. To efficiently optimize the ELBO, sampling from $q_\varphi(z|x)$ is done by reparametrizing $z$ as $z = \mu_\varphi(X) + \sigma_\varphi(X) \cdot \epsilon$, where $\epsilon \sim \mathcal{N}(0, \mathbb{1})$.

Since we approximate the prior $p_\theta(X)$ as locally normal distributed, the entropy is given by

\begin{align}
    \mathcal{H}(q_\varphi(z|X)) = \frac{d}{2}(1+ \ln(2\pi)) + \sum_{i=1}^{d} \ln \sigma_i.
\end{align}

To model a more complex latent distribution $p(z)$, the latent space $z$ is transformed by the \emph{LatentFlow}. The LatentFlow consists of multiple neural spline coupling layers \cite{Durkan:2019nsq} that transform the distribution to a normal distribution. Therefore, the expectation value of the prior distribution can be written as 

\begin{align}
\mathcal{L}_{\textrm{prior}} =   \mathbb{E}_{q_\varphi(z | X)}\left[
 \ln p(f^{-1}(z, \mathcal{C}))
+ \ln \left| \det \frac{\mathrm{d} f^{-1}(z, \mathcal{C})}{\mathrm{d} z} \right|\right].
\end{align}
The conditional probability density of the entire shower is approximated by the \emph{PointFlow}, which is implemented as a \deepsetflow, described in \cref{sec:deepsetflow}.

\begin{align}
    \mathcal{L}_{\text{recon}} = \mathbb{E}_{q_\varphi(z | X)}&\left[
    \ln p(g^{-1}(X, z, \mathcal{C})) + \ln \left| \det \frac{\mathrm{d} g^{-1}(X, z, \mathcal{C})}{\mathrm{d} X} \right|\right].
\end{align}

A schematic view of the sampling process is shown in \cref{fig:model-sampling}. For each shower, first $E_{\textrm{sum}} and n_{\textrm{hits}}$ are sampled by the \emph{CondFlow}. A latent vector $z$ is sampled by the \emph{LatentFlow}. At last, $n_{\textrm{hits}}$ points are generated by the \emph{PointFlow}.

\section{Pre- and Post-Processing}
\label{sec:prepostprocessing}

We transform the voxel based datasets to point cloud datasets. Each shower is represented by a combination of the coordinates ($(z, \alpha, r)$) and the energy ($e$) of all voxels for which $e > 0$.  To normalize the showers, we divide each energy value $e$ by the sum of all energies in the shower $E_{\mathrm{sum}}$. The energy fraction is min-max-scaled and transformed with the \emph{logit function}.

For the coordinates, we remove the $\alpha$-component, for further information, see \cref{sec:multiplehitkludge}. We dequantize $z$ and $r$ with the \cdfdequantization, described in \cref{sec:cdfdequantization}.

The conditional variables $c$ consists of the transformed number of hits $n_{\mathrm{hits}}$ and $E_{\mathrm{sum}}$
\begin{align}
c = \begin{pmatrix}
{\ln(n_{\mathrm{hits}} + \epsilon)}/{\sqrt{ E_{\mathrm{in}} }} \\
{E_{\mathrm{sum}}}/{E_{\mathrm{in}}}
\end{pmatrix}
\end{align}

Here $E_{\mathrm{in}}$ is the energy of the initial electron and the number of hits is dequantized by adding $\epsilon \sim U(0,1)$.

Both, $c$ and the transformed calorimeter hits are normalized to have a mean of 0 and a standard deviation of 1.

For sampling, we revert the transformation applied above to the generated points and conditional variables. The $\alpha$-component is reintroduced, as explained in \cref{sec:multiplehitkludge}.

\section{\cdfdequantization}

\label{sec:cdfdequantization}

Normalizing flows are invertible transformations utilized in machine learning to map  probability distributions to the normal distribution, primarily designed for continuous data. The application to discrete data, however, includes complexities.

Transforming discrete data into continuous ones is called \emph{dequantization}. For the new model we developed a novel method called \cdfdequantization, which utilizes the \emph{inverse transform method}, or \emph{Smirnov transform}, to map any  finite discrete one dimensional distribution to a normal distribution. We split the transformation in two parts. The first is a mapping to the uniform distribution between 0 and 1, here denoted as $U$. The second is a mapping from $U$ to the normal distribution $\mathcal{N}$.

Conventionally, dequantization involves the addition of uniformly distributed noise, as proposed by Uria et al. \cite{uria2014}, and subsequent scaling of discrete values to form a continuous-space dataset. The inverse of this process combines a back scaling operation with a floor operation. This method of dequantization assigns densities over hypercubes around data points, an essential step to prevent the density model from approximating a degenerate mixture of point masses.

However, modeling such hypercubes with smooth function approximators poses significant challenges. To overcome this, Ho et al. \cite{ho2019} introduced variational dequantization. Rather than assigning uniform noise, this technique incorporates an additional normalizing flow that learns and applies a structured, non-uniform noise to the data, effectively increasing the modeling complexity.

Dinh et al. \cite{dinh2017}, added another dimension to dequantization. They proposed applying a logit transformation to the dequantized variables, transforming the support from the interval [0, 1] to $(-\infty, \infty)$.

The logit transformation, when applied to $u \sim U$, maps to a logistic distribution. The challenge imposed to the normalizing flow is then to learn to convert this logistic distribution into a normal distribution, a task complicated by substantial differences in the distributions' tails. For further details, please refer to \cref{appendix-logistic-normal}.

Transforming a uniform distribution into any distribution causes a noteworthy problem in the domain of statistics and data science. One solution is using the inverse transform $\Phi_X^{-1}(u) = \inf\{x|F_X(x) \geq u\}$. This transform maps $U \to X$, where $X$ is any distribution. A proof for the case of a continuous distribution is provided in \cref{appendix-proof-continuous} or can be found in the statistics literature.

For our current focus, we are interested in mapping the standard uniform distribution onto the normal distribution, and also being able to reverse this process. To define these mappings, we rely on the cumulative distribution function (CDF) and its inverse, the quantile function. The CDF of a standard normal distribution, denoted $\mathcal{N}$, is given by

\begin{align}
    F_{\mathcal{N}}(x)=\frac{1}{2} \left[1 + \text{erf}\left( \frac{x}{\sqrt{2}} \right) \right] , 
\end{align}

where $\text{erf}(x)$ is the error function. This function provides the probability that a random variable from a standard normal distribution is less than or equal to $x$.  Conversely, the quantile function of the standard normal distribution, often referred to as the \textit{probit} function, is

\begin{align}
    F^{-1}_{\mathcal{N}}(y)= \sqrt{2} \text{erf}^{-1}(2y-1).
\end{align}

where $\text{erf}^{-1}(x)$ is the inverse error function. This function returns the value corresponding to a given probability $y$, such that the probability of a random variable from a standard normal distribution being less than or equal to this value is $y$. 

It is important to note that exact evaluations of the CDF and its inverse for the normal distribution are computationally expensive. However, approximations to these functions can be found in most numerical software libraries, making it efficient to perform the mapping between the uniform and normal distributions.

Let us extend our previous discussions to the scenario where we require a mapping from a given discrete probability distribution to the standard uniform distribution. To achieve this, we need to establish the validity of the inverse transform for discrete distributions. A proof is given in \cref{appendix-proof-discrete}.

\begin{figure}[H]
    \centering
    \includegraphics[width=\textwidth]{visualizations/cdf_dequantization.tikz}
    \caption{A schematic of the \cdfdequantization}
    \label{fig:cdfdequantization}
\end{figure}

The Smirnov transform gives us an approach to map the standard uniform distribution $U(0,1)$ to a discrete distribution by creating a decomposition of the interval $[0,1]$ into $n$ sub-intervals, each having a size equivalent to the corresponding probability $p_i$.

This forms a surjective mapping where intervals in $[0,1]$ are associated with specific discrete values. Notably, Nielsen et al. \cite{nielsen2020} illustrated how variational autoencoders (VAEs), normalizing flows, and surjective mappings can be integrated into one unified framework. They have shown that surjective mappings can be used if a sufficient stochastic inverse is found.

Within this framework, Nielsen et al. have shown that the act of adding uniform noise can be interpreted as a stochastic inverse of the floor operation, denoted as $\lfloor x \rfloor$. This implies that $p(x|z) = \delta_{x, \lfloor z \rfloor}$. Its stochastic inverse, $q(z|x)$, has support in $\mathcal{B}(x) = \{ x + u | u \in [0,1] \}$.

In the context of \cdfdequantization, the density $p(x_i | u) = \delta(F_X(x_i) \leq u \leq F_X(x_{i+1}))$ is a delta function that triggers if $u$ lies in the interval $[F_X(x_i), F_X(x_{i+1})]$. Its stochastic inverse, $q(u | x_i)$, is supported over 

\begin{align}
    \mathcal{B} &=\{u | F_X(x_i) \leq u \leq F_X(x_{i+1})\} \\
                &= \{F_X(x_i) + u | 0 \leq u \leq F_X(x_{i+1}) - F_X(x_i)\} \\
                &=\{F_X(x_i) + u | 0 \leq u \leq p_i\}\\
                &=  \{F_X(x_i) + u \cdot p_i | u \in [0,1] \}
\end{align}
For our mapping to be accurate, the distribution of $u$ values should be uniform across $[0,1]$. Therefore, we can compose the stochastic mapping as $\Phi_X(x_i) = F_X(x_i) + u \cdot p_i$ where $u \sim U(0,1)$.

If we have access to all $p_i$, this transformation provides a straightforward mapping from discrete data to a uniform distribution. The inverse transform can be found through a simple search across the probability intervals. We can simply approximate $p_i$ by counting the frequencies of $x_i$ in the data.

In summary, we have constructed a relatively simple transformation that can map discrete data to a normal distribution. In doing so, we have significantly eased the task of the normalizing flow, which is now left with only the responsibility to learn the correlations in the data and not the general shape. This enhances the efficiency and effectiveness of machine learning models dealing with discrete data. For a visual representation, see \cref{fig:cdfdequantization}
We provide the algorithms for both directions of the \cdfdequantization below.

\begin{minipage}{0.46\textwidth}
\begin{algorithm}[H]
    \centering
    \caption{Forward Transformation ($X \to \mathcal{N})$}\label{algorithm}
    \begin{algorithmic}
        \For{$x_i \in X$}
        \State sample $u \sim U(0,1)$
        \State $y_i = F_{\mathcal{N}}^{-1}(\text{CDF}(x_i) + \text{PDF}(x_i) \cdot u) $
        \EndFor
        \State \textbf{Return}  $Y = \{y_1 \dots y_n\}$
    \end{algorithmic}
\end{algorithm}
\end{minipage}
\hfill
\begin{minipage}{0.46\textwidth}
\begin{algorithm}[H]
    \centering
    \caption{Inverse Transformation ($\mathcal{N} \to X)$}\label{algorithm1}
    \begin{algorithmic}
    \For{$y_i \in Y$}
    \State $u_i = F_{\mathcal{N}}(y_i)$
    \State $x_i = \text{find first CDF} \geq u_i$
    \EndFor 
    \State \textbf{Return}  $X = \{x_1 \dots x_n\}$
    \end{algorithmic}
\end{algorithm}
\end{minipage}
\vspace{1em}

{\setlength{\parindent}{0cm}
This dequantization strategy is universal. For our \cpfii model we apply it to both the $r$ and $z$ dimensions of the points. The method treats each dimension independently and is thereby only altering the marginal distributions.}

\section{DeepSetFlow}

\label{sec:deepsetflow}

The previous version of \emph{PointFlow} utilized a system of coupling blocks to transform the features of points. This system divided the features into two equal parts, with one half undergoing transformation based on the other half. While permutation invariant and capable of handling varying cardinality, it lacked a mechanism for enabling information exchange among the points, resulting in an inability to model inter-point correlation.

To address this, we introduce a new \emph{Normalizing Flow} architecture called \deepsetflow. The architecture integrates \emph{DeepSets} \cite{deepsets} in each coupling layer. In the coupling layer, the part of all point features that is not further transformed is aggregated into a latent representation. The per-point transformation uses the latent representation as another conditional feature.

Graphically, our architecture resembles a central node that gathers information from all nodes. It distributes the information evenly and is linear in the number of points.

The new flow architecture is able to capture the point-to-point correlations.
This approach shares similarities with previous work by Buhmann et al. \cite{Buhmann:2023pmh} who constructed the EPiC-GAN, a model that applied \emph{DeepSets} to create a permutation equivariant generative model. Mikuni et al. \cite{Mikuni:2023dvk} also employed a similar technique of information aggregation within a diffusion model to learn point clouds.

In another instance, Käch et al. \cite{Kach:2023rqw} developed a model that combines all information into a single node, referred to as the mean field. Their model's main distinguishing factor from previous ones is the use of cross attention to update the information of the mean field.

Finally, Liu et al. \cite{liu2019} crafted a Graph Normalizing Flow. In creating a coupling layer for graphs, they suggested the same feature-splitting across points as in our model.

\section{Multiple Hit Workaround}

\label{sec:multiplehitkludge}

The dataset consists of voxels, thus only one hit per calorimeter cell is allowed. Point cloud-based models, however, generate points in continuous space and inherently have no limitation that only one point can exist within the space of a cell. Therefore, the model often produces multiple points per calorimeter cell, which is incompatible with the data. So far, no adequate solution to this problem is known and we have developed a novel approach to solve this problem.

We make use of the rotational symmetry of the calorimeter and generate the points without an $\alpha$-component. We distribute the points randomly in $\alpha$, and if there are more points than cells, these are randomly added to the occupied $\alpha$-positions. Here, we assume that there is no internal shower structure in $\alpha$. This is, of course, fundamentally incorrect. However, in our experiments, it improves the modelling of the electron showers. 

\section{Results}
\label{sec:results}

In this section, we assess the model's performance by comparing the generated showers with those simulated by \geant, using data that was not used in training. If not stated otherwise, we compare histograms of distributions in the figures below. The red dashed lines represent the original \geant data, while the blue lines represent the distribution produced by the \cpfii model. To provide a clearer perspective, smaller plots are included below each main figure, where the difference between the two distributions is expressed as a ratio.

\begin{figure}[H]
\centering
\begin{subcaptionblock}[b]{0.49\textwidth}
    \includegraphics[width=\linewidth]{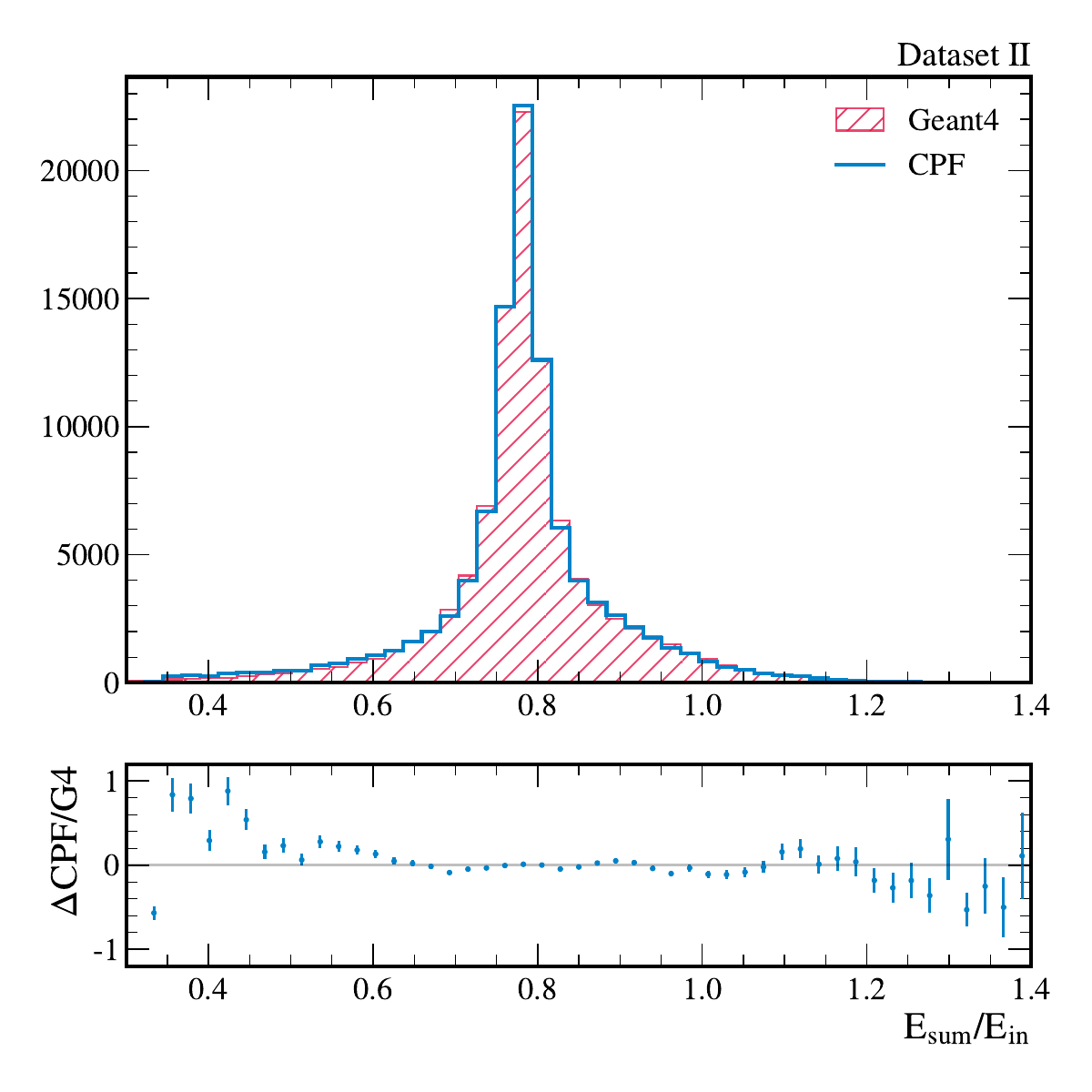} 
    \caption*{}
\end{subcaptionblock}
\begin{subcaptionblock}[b]{0.49\textwidth}
   \includegraphics[width=\linewidth]{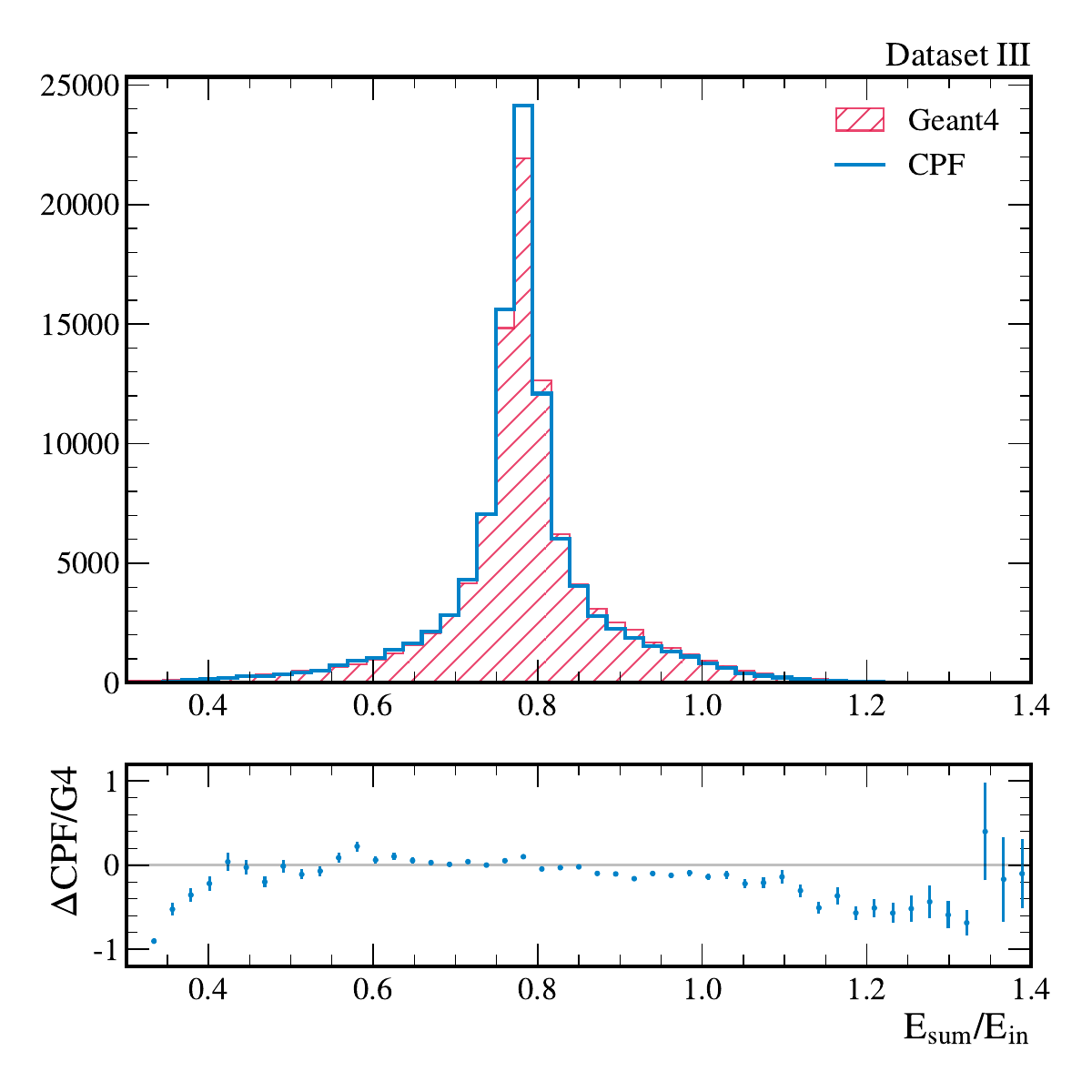} 
   \caption*{}
\end{subcaptionblock}
\caption{
The distribution of the sum of all energies in the calorimeter divided by the energy of the incident electron (left) for Dataset II and (right) for Dataset III.}
\label{fig:esum}
\end{figure}

\cref{fig:esum} shows the distribution of the sum of all energies in the calorimeter cells divided by the energy of the initial particle. A good agreement between \geant and \cpfii is concluded from this comparison.

\begin{figure}[H]
\centering
\begin{subcaptionblock}[b]{0.49\textwidth}
    \includegraphics[width=\linewidth]{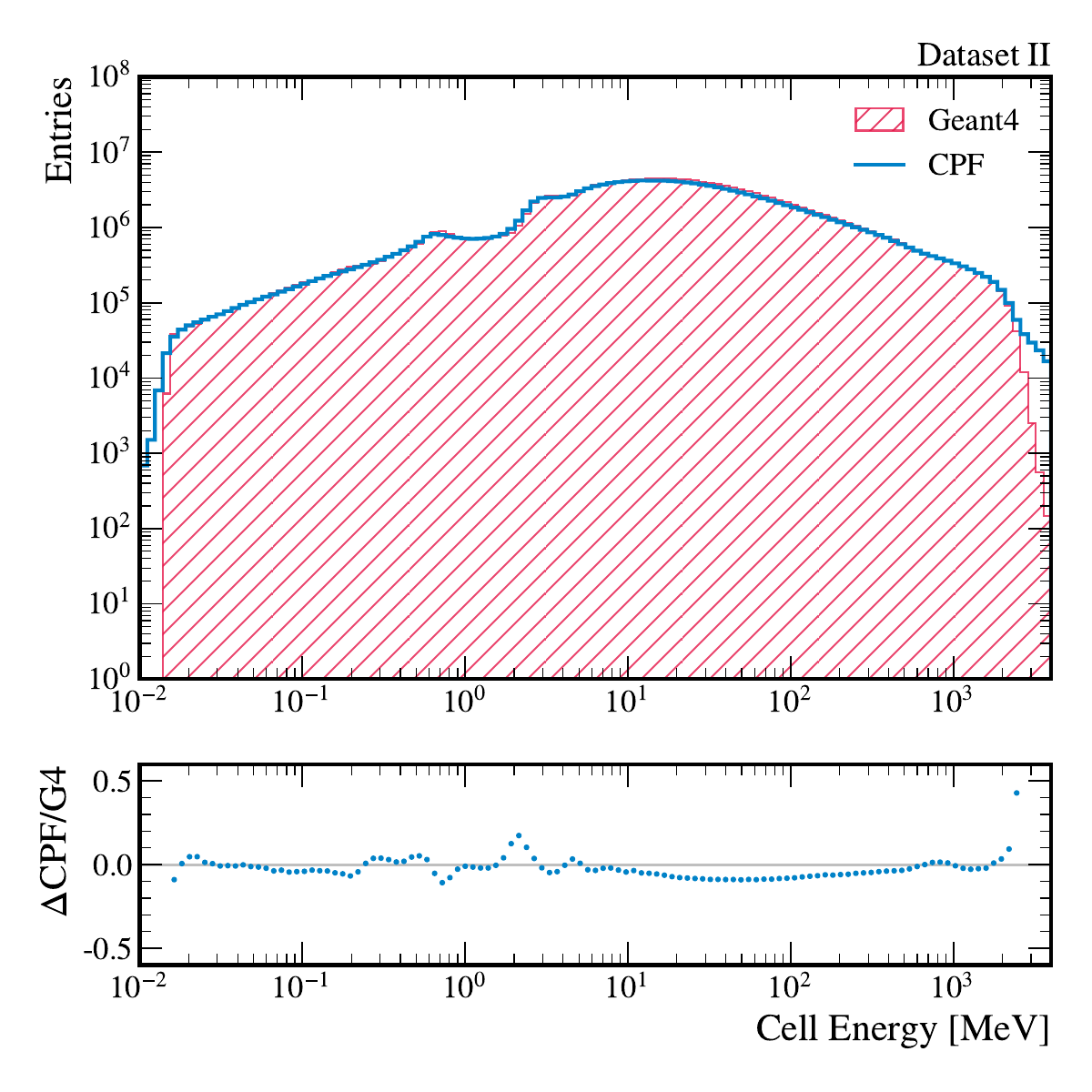} 
    \caption*{}
\end{subcaptionblock}
\begin{subcaptionblock}[b]{0.49\textwidth}
   \includegraphics[width=\linewidth]{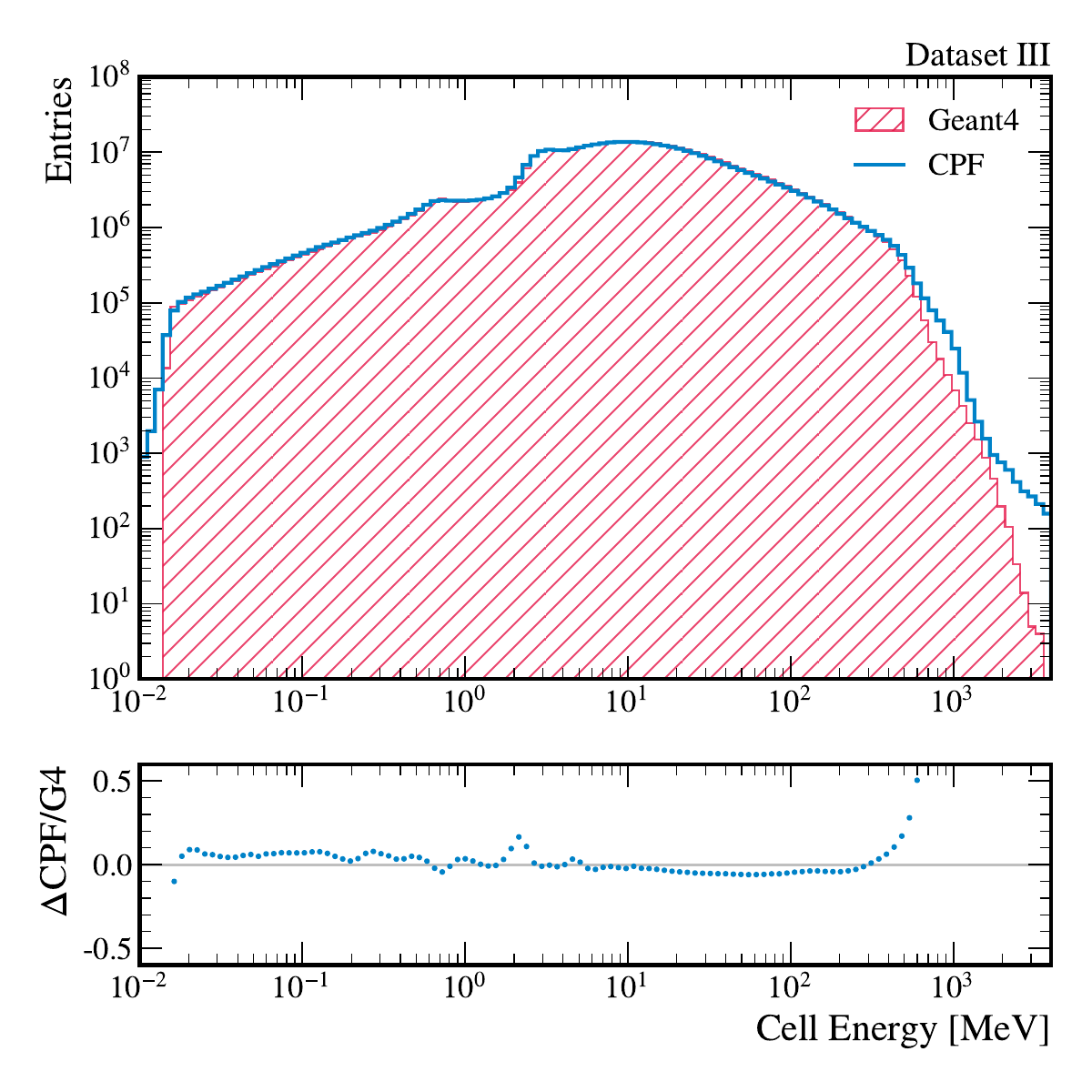} 
   \caption*{}
\end{subcaptionblock}
\caption{
The distribution of cell energies (left) for Dataset II and (right) for Dataset III.}
\label{fig:cell_energy}
\end{figure}

In \cref{fig:cell_energy} we present the energy distributions of individual cells in the two different datasets. These figures show consistency within the bulk of the distributions across both datasets. However, one observation can be made regarding the tails of these distributions. In both cases, the \cpfii model shows a tendency to overshoot or undershoot in energy. This phenomenon is particularly pronounced in Dataset III, where the deviations from the expected values are more pronounced in the tail regions.

\begin{figure}[H]
\centering
\begin{subcaptionblock}[b]{0.49\textwidth}
    \includegraphics[width=\linewidth]{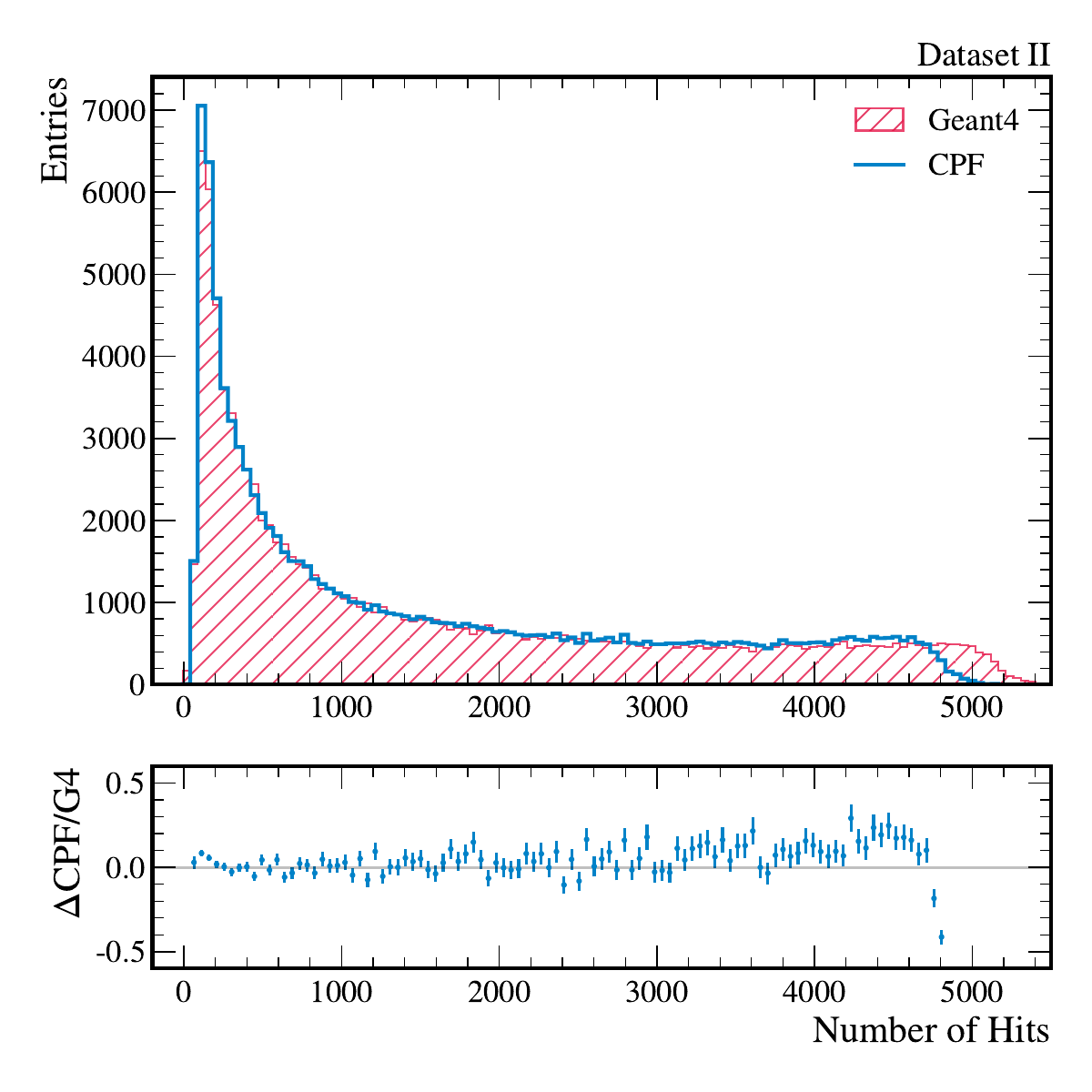} 
    \caption*{}
\end{subcaptionblock}
\begin{subcaptionblock}[b]{0.49\textwidth}
   \includegraphics[width=\linewidth]{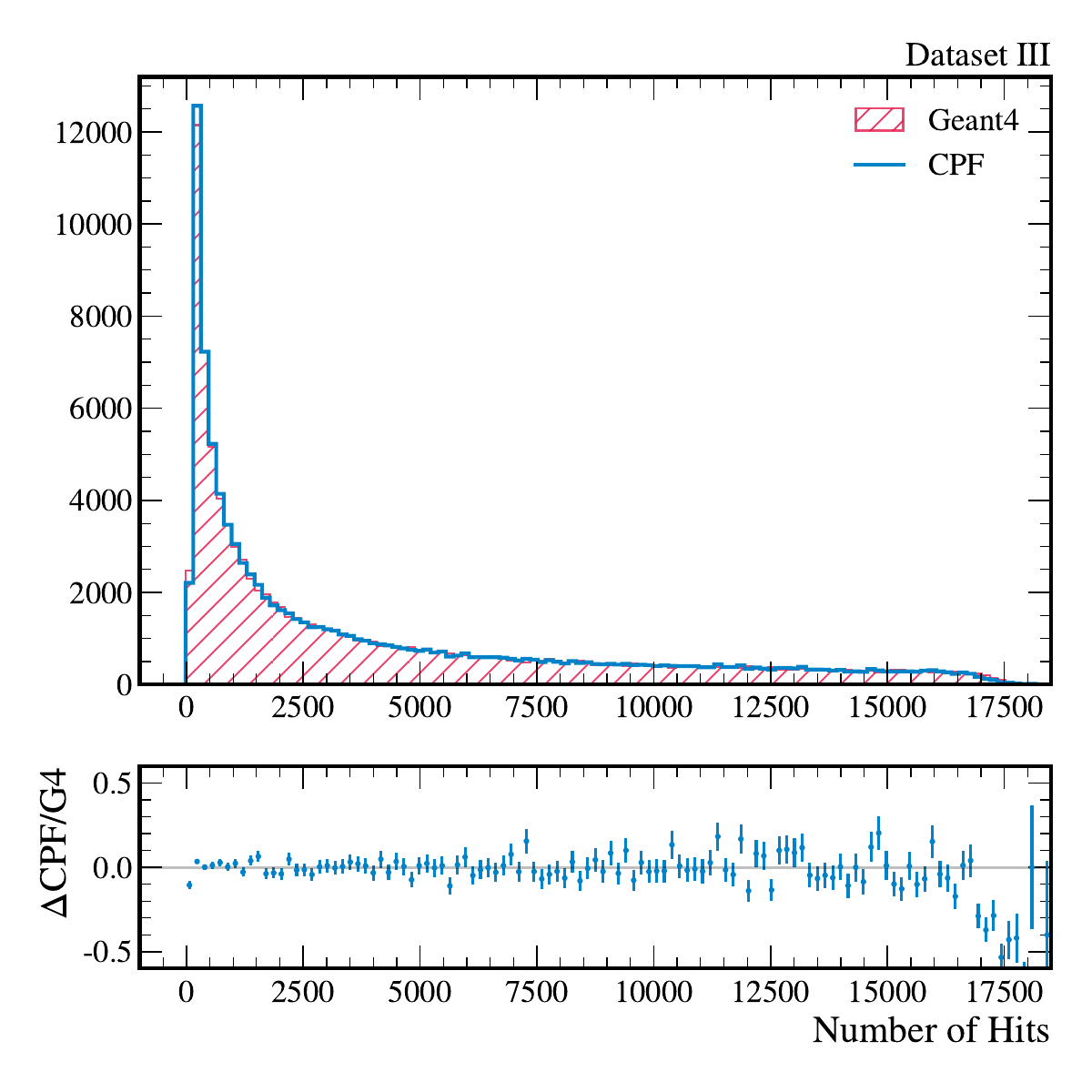} 
   \caption*{}
\end{subcaptionblock}
\caption{
The distribution of number of hits (left) for Dataset II and (right) for Dataset III.}
\label{fig:num_hits}
\end{figure}

\cref{fig:num_hits} shows the distribution of the number of hits in both data sets. The "number of hits" is defined as the number of calorimeter cells that have an energy value greater than zero for a given shower. A key observation in both datasets is the generally accurate modeling of the overall distribution of hits.  However, a closer examination of \cref{fig:num_hits} reveals a limitation in the models ability to reproduce the highest number of hits. This problem is most likely due to the multiple hits workaround. In scenarios where most detector cells are hit, the probability of mapping multiple points to a single cell increases significantly. This results in a discrepancy in the tail of the distribution, where the model fails to generate the extreme values observed in the real data.
Conversely, for Dataset III this particular problem is not apparent. This dataset is characterized by a smaller proportion of total cells being hit. This observation suggests that the problem of accurately modeling the highest number of hits is less pronounced in more granular calorimeters. 

\begin{figure}[H]
\centering
\begin{subcaptionblock}[b]{0.9\textwidth}
    \includegraphics[width=\linewidth]{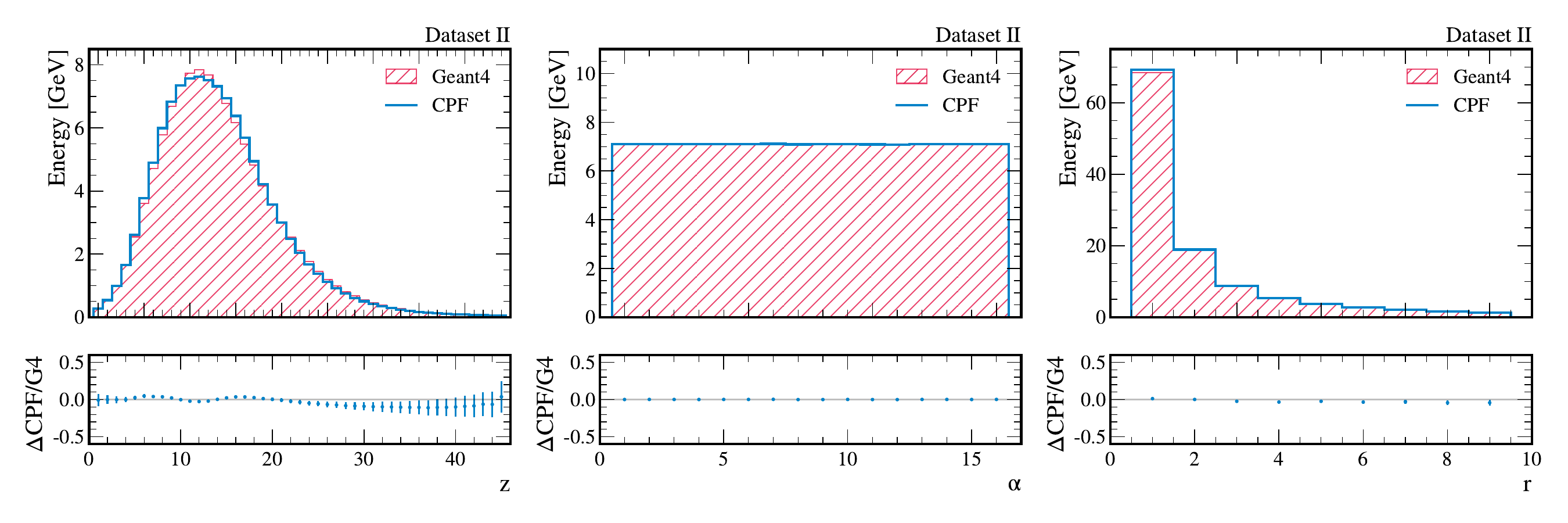} 
    \caption*{}
\end{subcaptionblock}

\begin{subcaptionblock}[b]{0.9\textwidth}
    \includegraphics[width=\linewidth]{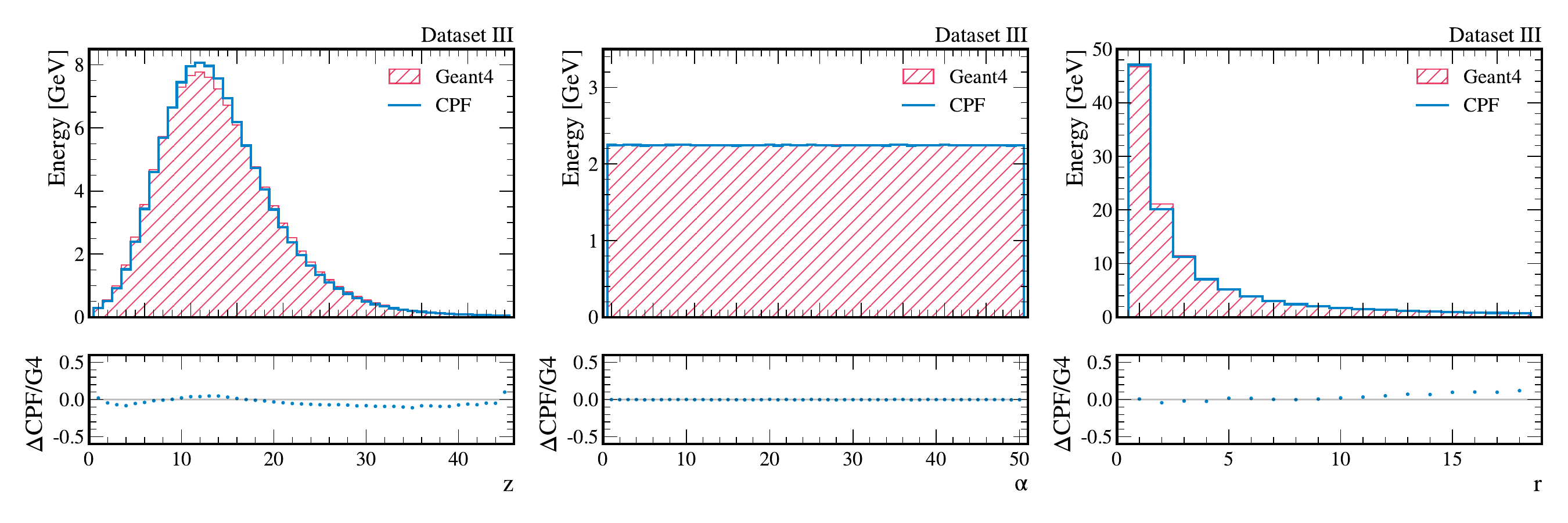} 
    \caption*{}
\end{subcaptionblock}
\caption{
The marginal profiles of the average shower shapes in $z, \alpha$ and $r$ (top) for Dataset II and (bottom) for Dataset III.}
\label{fig:res_shower_shape}
\end{figure}

\cref{fig:res_shower_shape} shows the marginal profiles of the average shower for the second and third data sets, respectively. \cref{fig:res_shower_shape} (left) shows the longitudinal shower profile. We observe minimal variation between the two distributions, indicating a strong agreement in the longitudinal profile.
In \cref{fig:res_shower_shape} (middle) the flat distribution of $\alpha$ is visible. Both distributions appear to be uniformly flat, which is an inherent property of the construction. Finally, \cref{fig:res_shower_shape} (right) both figures examines the radial distribution, where we find that both distributions have similar decreasing structure.

\begin{figure}[H]
\centering
\begin{subcaptionblock}[b]{0.9\textwidth}
    \includegraphics[width=\textwidth]{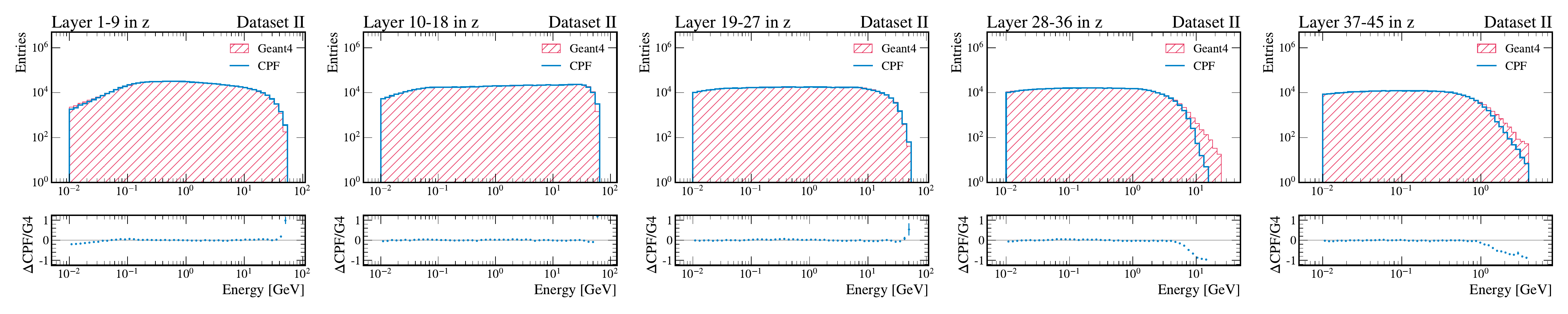}
    \caption*{}
\end{subcaptionblock}

\begin{subcaptionblock}[b]{0.9\textwidth}
    \includegraphics[width=\textwidth]{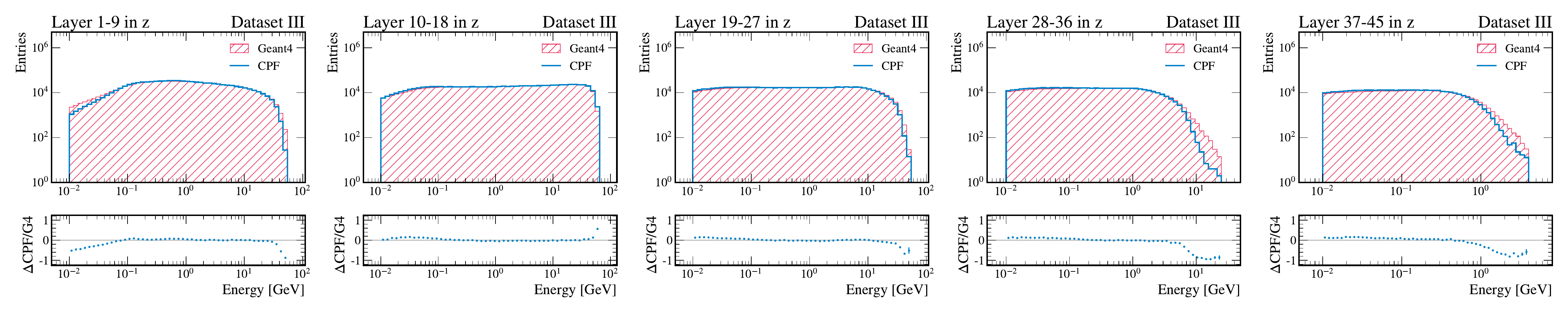} 
    \caption*{}
\end{subcaptionblock}
\caption{
The energy distribution in different layers of the detector (top) for Dataset II and (bottom) for Dataset III.}
\label{fig:energy_layer}
\end{figure}

\cref{fig:energy_layer} provides a detailed view of the distribution of the total energy in different areas of the detector. For each dataset,  five plots are shown with the energy distribution in five successive layers of the detector. A consistent pattern observed in these figures is the accurate modeling of the energy distributions in the initial layers of the detector. However, there is a notable discrepancy in the rear layers. In these layers, there is a significant absence of higher energy values in the model results. 

\begin{figure}[H]
\centering
\begin{subcaptionblock}[b]{0.9\textwidth}
    \includegraphics[width=\textwidth]{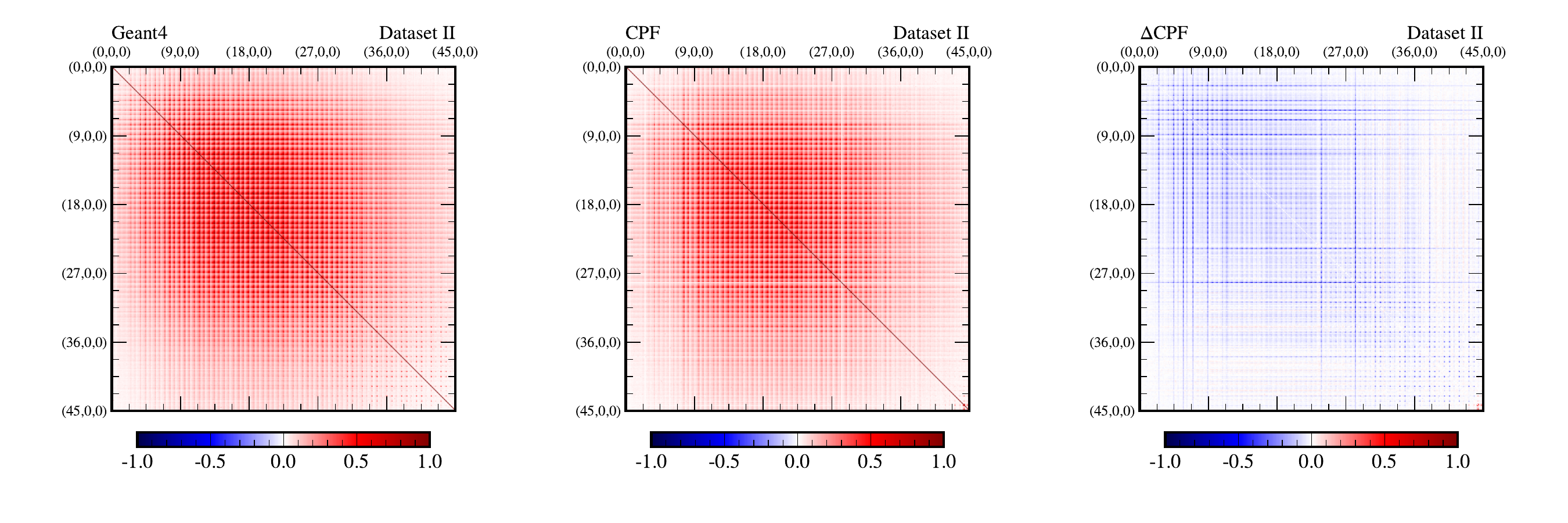}
    \caption*{}
\end{subcaptionblock}

\begin{subcaptionblock}[b]{0.9\textwidth}
    \includegraphics[width=\textwidth]{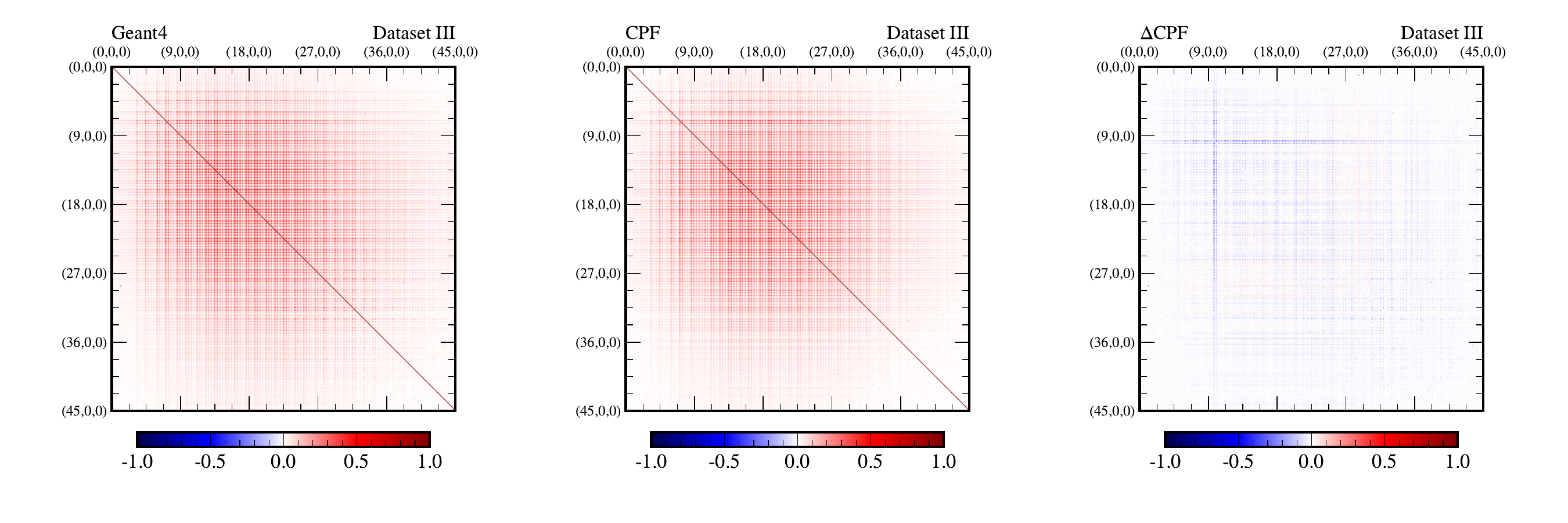} 
    \caption*{}
\end{subcaptionblock}
\caption{
The Pearson correlation coefficients between all cells of the detector are displayed (top) for Dataset II and (bottom) for Dataset III.}
\label{fig:corrcoef}
\end{figure}

\cref{fig:corrcoef}  contains the Pearson correlation coefficients, which quantify the relationships between the energies of all cells in the detector. For visualization purposes, the three-dimensional structure of the detector is flattened, resulting in a recurring pattern within the displayed data. \cref{fig:corrcoef} (left) shows the energy distribution as simulated by \geant, \cref{fig:corrcoef} (middle) illustrates the results from the \cpfii model, and \cref{fig:corrcoef} (right) highlights the differences between the two models. A key observation is the overall effective modeling of the correlation coefficients by the \cpfii model. This suggests that the model is able to capture the inter-cell energy relationships within the detector.

\begin{figure}[H]
\centering
\begin{subcaptionblock}[b]{0.9\textwidth}
    \includegraphics[width=\textwidth]{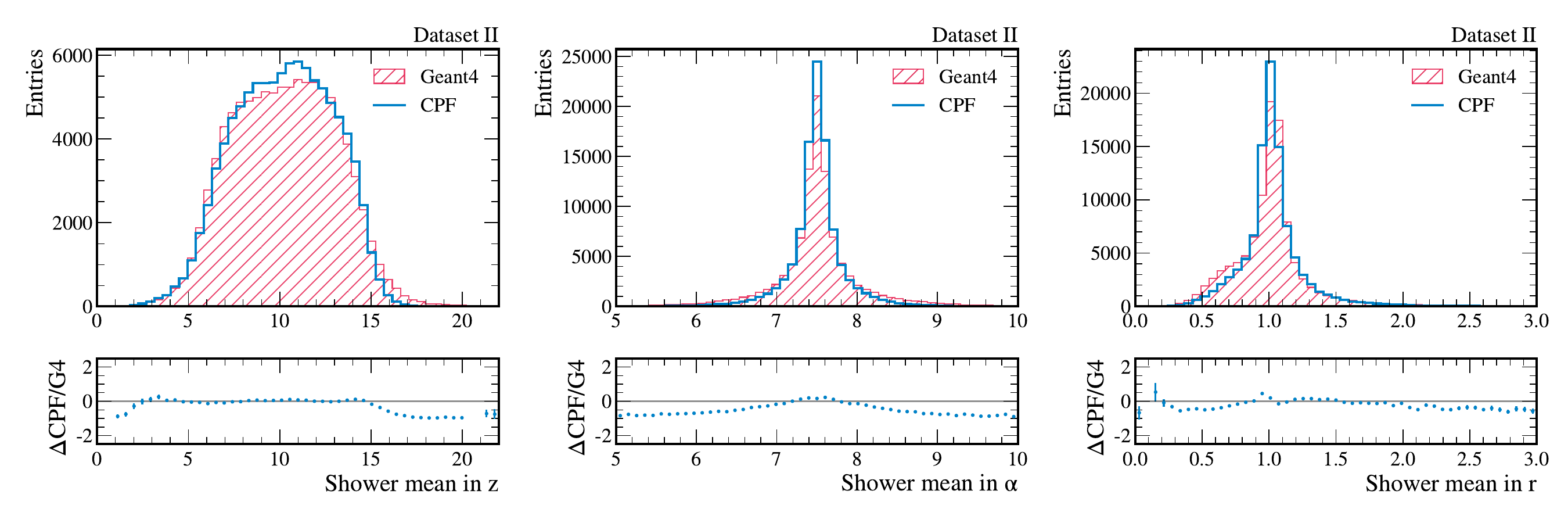}
    \caption*{}
\end{subcaptionblock}

\begin{subcaptionblock}[b]{0.9\textwidth}
    \includegraphics[width=\textwidth]{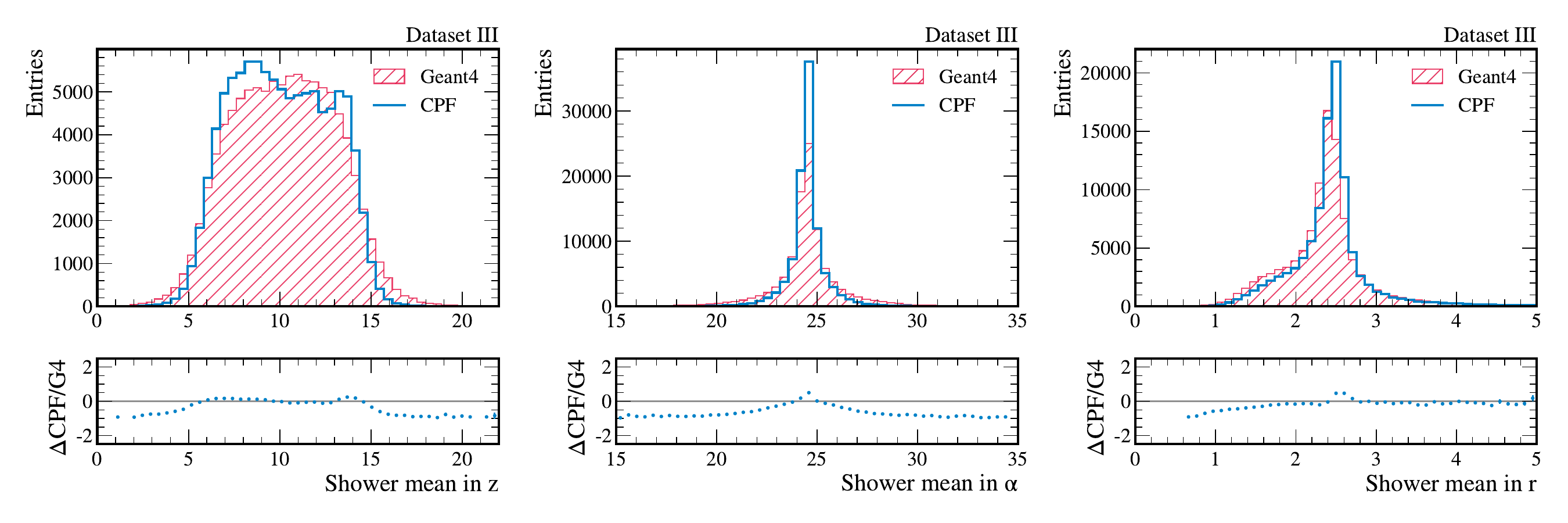} 
    \caption*{}
\end{subcaptionblock}
\caption{
Energy-weighted shower center distribution in all three dimensions (top) for Dataset II and (bottom) for Dataset III.}
\label{fig:shower_means}
\end{figure}

\cref{fig:shower_means} shows the energy-weighted mean of the shower in Dataset II and III, respectively.  The means $\vb{\mu}_{i}$ are calculated as 
\begin{align}
\vb \mu_i = \frac{\sum_{j} \vb c^i_j \cdot e^i_{j}}{\sum_{j} e^i_{j}},
\end{align}
where $\vb c^i_{j}$ are the 3-dimensional coordinate vector of the hit constituents of the shower and $e_{j}^i$ are the corresponding energies.
The modeled showers capture the general trend of the energy-weighted mean, but there are some discrepancies with the actual data. These differences may indicate limitations in the model's ability to accurately simulate the nuanced distribution of energy within the showers.

\begin{figure}[H]
\centering
\begin{subcaptionblock}[b]{0.9\textwidth}
    \includegraphics[width=\textwidth]{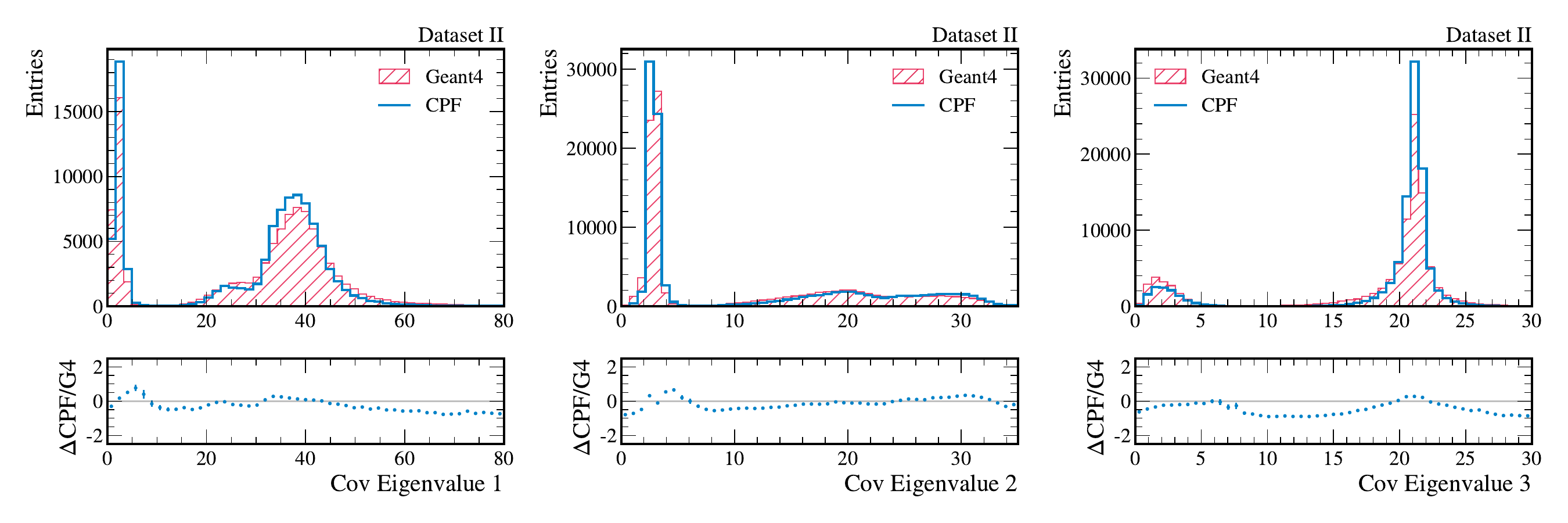}
    \caption*{}
\end{subcaptionblock}

\begin{subcaptionblock}[b]{0.9\textwidth}
    \includegraphics[width=\textwidth]{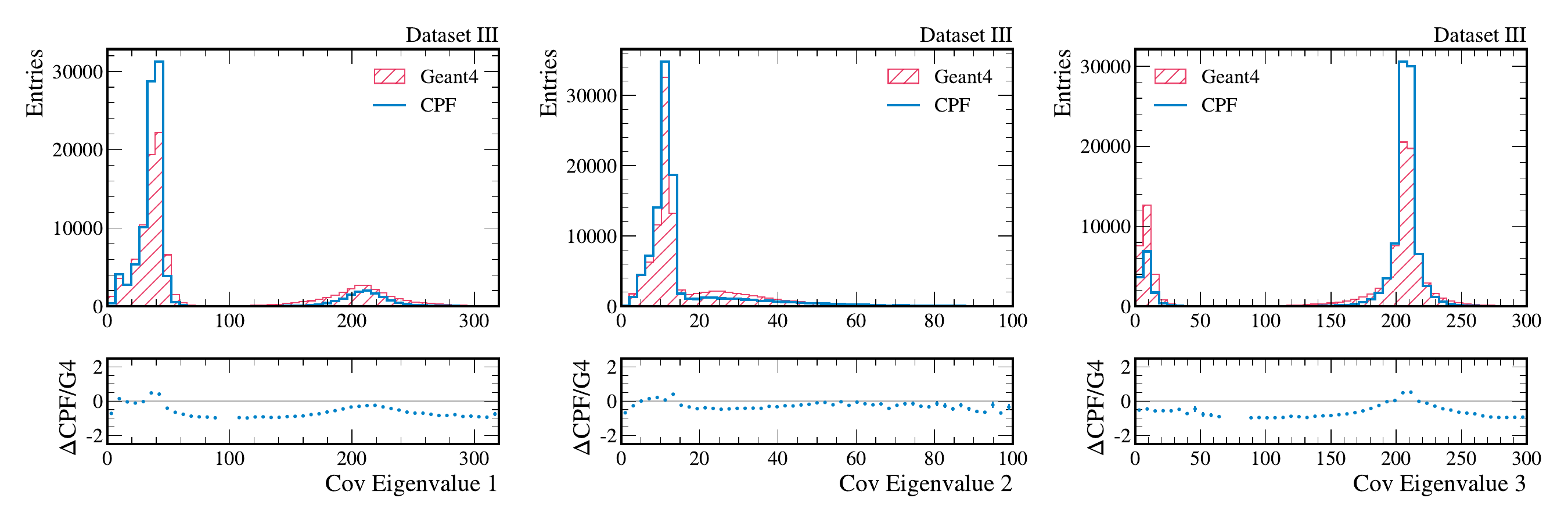} 
    \caption*{}
\end{subcaptionblock}
\caption{
The eigenvalues of the covariance matrix weighted by energy are presented (top) for Dataset II and (bottom) for Dataset III.}
\label{fig:cov_eigen}
\end{figure}

To further explore the structure of the showers in \cref{fig:cov_eigen}, we compute the energy-weighted covariance matrix for each shower. This is done by first determining the deviation of each cell position vector $\vb c_j^i$ from the mean position vector $\vb \mu_i$ of the shower. This deviation is denoted as $\hat{\vb{c}}_{j}^i = \vb c_j^i - \vb \mu_i$. The energy-weighted covariance matrix $C_{ik}$ is then computed using the formula

\begin{align}
C_{ik} = \frac{\sum_{j} e^i_{j} (\hat{\vb{c}}_{j}^i)^T \cdot \hat{\vb{c}}_{j}^i}{\sum_{j} e^i_{j} - 1}.
\end{align}

This matrix represents the spread and orientation of the shower in the detector space, energy-weighted. After computing these matrices, we decompose them to extract their eigenvalues. These eigenvalues, which represent the principal components of the showers spread in three dimensions, are then plotted to analyze their distribution.

In analyzing these distributions, we observe a notable contrast between the two datasets. For Dataset II, there is good agreement between the model and the actual data, indicating that the model effectively captures the spatial structure of the showers in this dataset. For Dataset III, however, the discrepancies are more pronounced. 

\begin{table}
    \centering
    \begin{tabular}{|c|l|cc|cc|}
    \hline
    \multirow{2}{*}{Dataset} & \multirow{2}{*}{Model} & \multicolumn{2}{|c|}{low level classifier} & \multicolumn{2}{|c|}{high level classifier}  \\
     &  & AUC & JSD & AUC &  JSD \\
    \hline
    \multirow{2}{*}{II} & CPF I & $0.945 \pm 0.004$ & $0.604 \pm 0.014$ & $0.927 \pm 0.003$ & $0.509 \pm 0.008$ \\
     & CPF II & $0.826 \pm 0.006$ & $0.275 \pm 0.008$ & $0.785 \pm 0.009$ & $0.220 \pm 0.015$ \\
    \hline 
    \multirow{2}{*}{III} & CPF I & $0.786 \pm 0.019$ & $0.121 \pm 0.022$ & $0.947 \pm 0.003$ & $0.582 \pm 0.010$ \\
     & CPF II & $0.709 \pm 0.040$ & $0.107 \pm 0.029$ & $0.934 \pm 0.003$ & $0.530 \pm 0.008$ \\
    \hline
    \end{tabular}
    \caption{\challenge Classifier Score for the \cpf I and \cpfii model.}
    \label{tab:classifier_score}
\end{table}

The performance of the official \challenge classifier is evaluated in  \cref{tab:classifier_score}. The classifier is applied ten times to each dataset and model. The evaluation shows that the \cpfii model outperforms the \cpf I model in both low-level and high-level classifications across all datasets. This significant difference in performance highlights the advancements and improvements incorporated in the \cpfii model.
For Dataset III, there is a noticeable trend where the low-level classifier is less effective for distinguishing between outcomes compared to the high-level classifier.
It is possible that further tuning or retraining of the model, with a focus on its handling of high-level features, could lead to further improvements.

\section{Conclusion}

\label{sec:conclusion}

In this study, we have introduced \cpfii, a refined generative model that significantly advances the simulation of calorimeter showers. The utilization of point clouds, owing to their inherent sparsity, offers a computationally more efficient approach compared to traditional fixed data structures like 3D images. Building upon the foundations of the original \cpf model, \cpfii incorporates innovative techniques such as a novel dequantization method, referred to as \cdfdequantization, and a new normalizing flow architecture, named \deepsetflow. These enhancements contribute to a more efficient and accurate modeling of particle interactions in calorimeters, enabling more efficient and accurate modeling of complex particle interactions. 

The extensive evaluation of \cpfii using the challenging \challenge datasets II and III has proven its efficacy. Our model not only exhibits superior performance in both low-level and high-level classifications compared to its predecessor but also showcases significant improvements in the accuracy of calorimeter shower simulations. The results demonstrate not only its capability to accurately simulate calorimeter showers, but also a marked improvement in computational efficiency compared to traditional methods. The model exhibits a balance between fidelity in the simulation and the computational demands, especially evident in the precise modeling of the spatial structure and energy distributions within calorimeter showers.

In comparison to other models, \cpfii shows notable advancements, particularly in terms of computational speed and accuracy. While there remain areas for further improvement, particularly in aligning the model's output more closely with real-world data, the strides made with \cpfii are clear. Future work could involve exploring more sophisticated architectural designs for the model to enhance its performance further.

Furthermore, the upcoming \challenge summary paper will provide a comprehensive comparison of \cpfii against other models.

\section{Acknowledgements}
The Deutsches Elektronen-Synchrotron DESY, a member of the Helmholtz Association (HGF), funded this research. The Maxwell computing resources operated at DESY were used to perform this research.

\clearpage

\bibliography{literature}

\appendix

\section{Logistic Distribution and Normal Distribution}
\label{appendix-logistic-normal}

\begin{figure}
    \centering
    \includegraphics[width=0.8\linewidth]{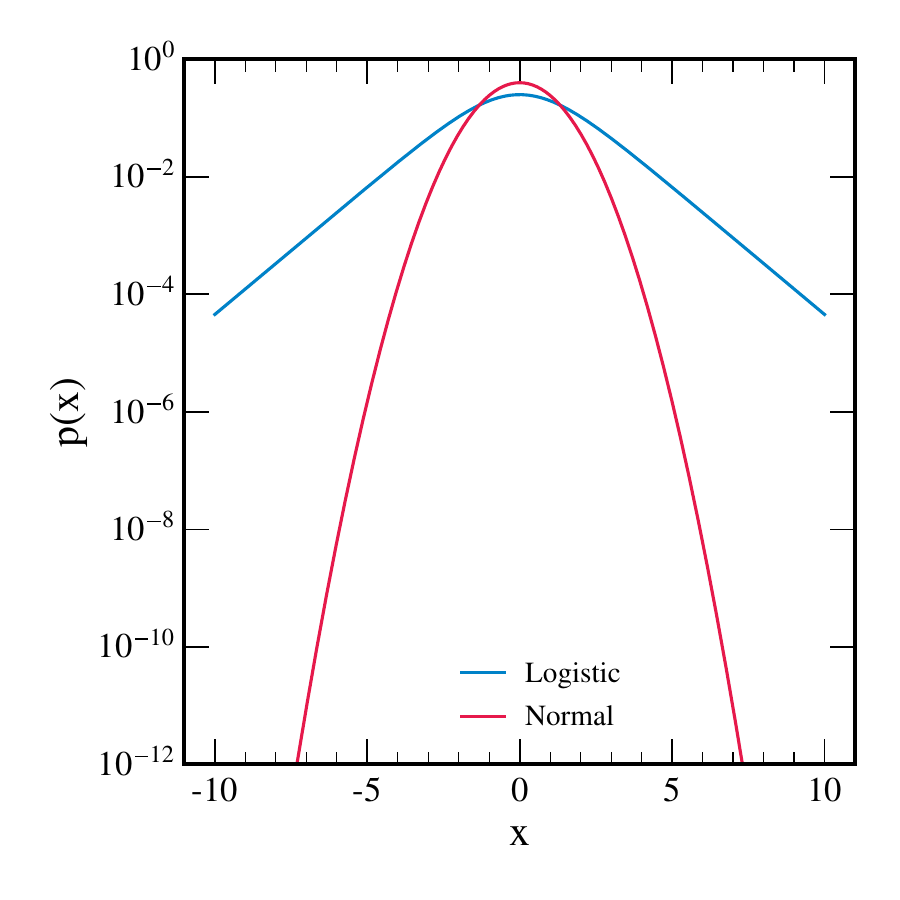}
    \caption{Comparison of the logarithmic declines of the logistic and normal distribution.}
    \label{fig:app_logistic_normal}
\end{figure}

\cref{fig:app_logistic_normal} shows that the tails of the normal distribution decrease significantly faster than the logistic distribution.
\section{Proof Inverse Transformation for continuous distributions}
\label{appendix-proof-continuous}

Let's consider $X$ as a random variable that has a cumulative distribution function (CDF) represented as $F_X(x) = P(X \leq x)$. The CDF of a random variable is an essential element in probability and statistics, as it provides the probability that a random variable will take a value less than or equal to a specific value.

By the properties of the CDF, we know that if $x < y$, then $F_X(x) \leq F_X(y)$, because the probability that $X$ is less than or equal to $x$ is always less than or equal to the probability that $X$ is less than or equal to $y$. Moreover, $F_X$ is strictly increasing where the probability density function $P_X(x) > 0$, indicating that the probability increases as the value of $X$ increases.

Now, let's suppose that $F_X$ is strictly increasing. Then, for any $u \in (0,1)$, the equation $F(x) = u$ has exactly one solution. We denote this solution as $x = F^{-1}_X(u)$, where $F^{-1}_X$ is the inverse function of $F_X$. In such a situation, we can say that  

\begin{align}
\Phi_X^{-1}(u)= \inf\{x|F_X(x) \geq u\}= \inf\{x|F_X(x) = u\} = F_X^{-1}(u)
\end{align}

This shows that the Smirnov transformation is essentially the inverse of the CDF, provided $F_X$ is strictly increasing.

Furthermore, the inverse function $F^{-1}_X$ is also strictly increasing on the interval $(0,1)$. This is because if $F_X$ is strictly increasing, then the inverse function will also preserve this property.

Let's now define a new random variable $Y = F_X^{-1}(U)$. For this random variable, we can express the cumulative distribution function of $U$ as $F_U(F_X(x))=P(U \leq F_X(x)) = F_X(x)$.

Given that $F^{-1}_X$ is strictly increasing, we can proceed and apply this property to our inequality. Specifically, $P(U \leq F_X(x)) = P(F_X^{-1}(U) \leq F_X^{-1}(F_X(x)))$. Since $F_X^{-1}(F_X(x))$ simplifies to $x$, this can be re-written as $P(F_X^{-1}(U) \leq x) = P(Y \leq x) = F_Y(x)$. Hence, we have derived that $F_Y(x) = F_X(x)$.

In conclusion, given these results and by the definition of the equality of random variables, we can say that $X = Y$.

As we've established, the equality $X = F_X^{-1}(U)$ and its inverse $U = F^{-1}_X(X)$ provide an invertible mapping between the standard uniform distribution and any continuous distribution without gaps. 

\section{Proof Inverse Transformation for discrete distributions}

\label{appendix-proof-discrete}

Suppose $X$ is a discrete random variable with $p_i = P(X = x_i)$ for $i \in {1, ..., n}$, where $p_i$ is the probability that $X$ equals $x_i$ and $n$ is the total number of possible outcomes. The cumulative distribution function $F_X(x_n)$ is given by $\sum_i^n p_i$.

Let's consider an arbitrary interval $[a, b]$ such that $0 \leq a \leq b \leq 1$. In the standard uniform distribution, the probability that $U$ falls within this interval is

\begin{align}
P(a \leq U \leq b) = P(U \leq b) - P(U \leq a) = b - a.
\end{align}

Now, for all indices $i < j$, we have $0 \leq F_X(x_i) \leq F_X(x_{j}) \leq 1$ due to the properties of the cumulative distribution function. Consequently, the probability that $U$ lies between $F_X(x_i)$ and $F_X(x_{i+1})$ equals $P(F_X(x_i) \leq U \leq F_X(x_{i+1})) = F_X(x_{i+1}) - F_X(x_i) = p_{i+1}$. This probability is precisely the probability that the discrete random variable $X$ equals $x_{i+1}$.

Based on these results, we can construct the following function that acts as the inverse transform for the discrete distribution

\begin{align}
      \Phi^{-1}_X(u) = 
  \begin{cases}
    x_1, & \text{if } U\leq F_X(x_1) \\
    x_2, & \text{if } F_X(x_1)\leq  U\leq F_X(x_2)\\
    \vdots \\
    x_n, & \text{if } F_X(x_{n-1})\leq  U\leq F_X(x_{n})\\
   \end{cases}.
\end{align}

In other words, $\Phi^{-1}_X(u)$ assigns the value $x_i$ to $u$ if it falls within the interval $[F_X(x_{i-1}), F_X(x_i)]$.
With this piece-wise defined function, we can express the inverse transform as 

\begin{align}
\Phi^{-1}_X(u) = \inf\{x_i | u \leq F_X(x_i)\}
\end{align}

which provides us with the required mapping from the standard uniform distribution $U$ to the discrete random variable $X$. Hence, we have successfully demonstrated the inverse transform method.

\end{document}